\documentclass[twocolumn,showpacs,amsmath,amssymb,aps,prb,floatfix]{revtex4}
\usepackage{graphicx}
\usepackage{dcolumn}
\usepackage{bm}

\begin{document}

\title{Calculation of hysteretic force between a superconductor and a magnet}

\author{M. J. Qin}
\author{G. Li}
\author{H. K. Liu}
\author{S. X. Dou}
\affiliation{Institute for Superconducting and Electronic Materials, University of Wollongong, Wollongong NSW 2522, Australia}
\author{E. H. Brandt}
\affiliation{Max-Planck-Institut f\"ur Metallforschung, Institut f\"ur Physik, D-70506 Stuttgart, Germany}

\date{\today}

\begin{abstract}
The magnetic levitation forces of a superconducting disk (SC) levitated by a cylindrical permanent magnet (PM) have been calculated from first principles for superconductors with finite thickness. The current $j(\rho,z)$ and field $B(\rho,z)$ profiles in the SC in the non-uniform magnetic field generated by the PM have been derived. The levitation force is observed to depend non-linearly on the critical current density $j_c$ and on the thickness of the SC. The flux creep is described by a current-voltage law $E(j)=E_c(j/j_c)^n$, from which we show that the levitation force depends on the speed at which the PM approaches or recedes from the SC, which accounts for the experimentally observed force creep phenomenon. The stiffness of the system has been derived by calculating the minor force loops. The numerical results reproduce many of the features exhibited by experiments.
\end{abstract}
\pacs{74.60.-w, 74.25.Ha, 74.25.Ld}
\maketitle

\section{\label{sec:intr}Introduction}

It has been well known that a permanent magnet (PM) can be stably
levitated above a high temperature superconductor (HTS) cooled by
liquid nitrogen, which has become the well known symbol for HTS
technology. This fascinating magnetic levitation results from the
interaction of the induced current inside the superconductor with
the inhomogeneous magnetic field generated by the PM. Because of
its possible industrial applications, such as noncontacted
superconducting bearings \cite{hull95,weinberger91}, gravimeters
\cite{goodkind99}, flywheel energy storage systems
\cite{hull00,hull94,kameno99,bornemann94,chen94,coombs98},
magnetic levitation transport systems \cite{tosa98}, and motors
\cite{oswald99}, the magnetic levitation between a PM and a HTS
has been the subject of intensive studies for the last decade.

The most common feature of the magnetic levitation is the
hysteretic behavior of the vertical force $F_{\rm z}$ versus the
distance $z$ between the PM and the HTS when the PM is descending
to and then ascending from the HTS. When a PM approaches a
zero-field-cooled HTS, the levitation force increases
monotonically from zero, as the PM is moving from the HTS, the
levitation force decreases sharply to a negative peak at some
distance, indicating attractive force between the HTS and the PM,
then declines to zero again at larger distance. Detailed
experiments have been performed on melt-textured-grown (MTG) YBCO
and RBCO (R denotes a rare earth element such as Nd, Sm, Gd, Eu,
Dy, Ho, Er, Tm, Yb, Lu and La), thin films, as well as granular
samples \cite{hull00,hull99,ullrich96,shi97,hennig98}. For MTG
YBCO and RBCO samples, the force curve is usually asymmetrical,
i.e., the absolute value of the attractive force is smaller than
the maximum repulsive force, $|F_{\rm a,max}|<F_{\rm r,max}$. For
thin films, the force curve is almost symmetrical $|F_{\rm
a,max}|\approx F_{\rm r,max}$ \cite{riise99}, while for granular
samples, the attractive force is hardly present\cite{chang90}.

Various parameters are very important in determining the
levitation force. The most important are the pinning strength
(critical current density) of the HTS and the induced shielding
current loops inside the HTS. Higher $j_c$ and larger loops are
very important to achieve a high levitation force. It has been
suggested that the levitation force increases linearly with $j_c$
and more than linearly with the current loops \cite{ullrich96}.
According to this criterion, the current material of choice for
superconducting levitation is MTG YBCO and RBCO samples, because
high quality MTG samples with high critical current density and
single domain diameters up to 10 cm are currently available by
means of melt process \cite{murakami92}. Another reason is its
high irreversibility line at liquid nitrogen temperature, which is
the normal operating temperature for the magnetic levitation
experiments. Although higher critical current density can be
achieved in thin films, the levitation force is limited by the
thickness of the films.

According to the critical state model \cite{bean62,bean64}, the
levitation force is independent of the speed at which the PM
approaches and recedes from the HTS \cite{moon94}. However, as the
thermally activated flux motion is prominent in HTS, resulting in
the relaxation of the magnetization (current density),
\cite{yeshurun88} the levitation force which depends on the
current density of the HTS is expected to decrease with time
(force creep). Experimentally the levitation force is observed to
be approximately logarithmic in time and can be well correlated
with the thermally activated flux motion in HTS \cite{riise92}.

The thickness of the HTS also drastically influences the
levitation force. Because the critical current density is limited,
for thin HTS samples, the levitation force increases linearly with
the thickness of the HTS. However, beyond a certain thickness, the
levitation force is independent of the thickness
\cite{unsworth93}. Other characteristics of the HTS, such as the
anisotropy and the grain orientation inside the MTG samples have
been studied and shown to affect the levitation force
\cite{shi97,tent98}. The stiffness of the PM-HTS system, which
represents the spring constant associated with vibrational motion
of the levitation system has been intensively studied. And
Experimental results for vertical stiffness, lateral stiffness and
cross stiffness have been presented \cite{hull00,hull99}.

Besides these, the geometry and condition of the PM also affect
the levitation force. The size and shape of the PM, the
homogeneity and the temperature dependence of the PM magnetization
have been shown to influence the levitation force \cite{hull99}.

Although the magnetic levitation between a PM and a HTS can be
easily demonstrated and detailed experimental results for the
levitation force have been presented, the theoretical models for
this magnetic levitation system have not been fully developed yet
so far. The first reason is that a model has to consider a finite
thickness superconducting disk under a perpendicular magnetic
field. In the case of an infinite long cylinder under a parallel
applied magnetic field, the magnetic field and current density
profiles can be easily obtained by means of the critical state
model. However, for a superconducting disk under a perpendicularly
applied magnetic field, the extreme demagnetization effects make
this case qualitatively different from the parallel field case.
The second reason is that the model has to consider the response
of a HTS immersed in a non-uniform magnetic field generated by the
PM.

In order to avoid the above difficulties, all the models for the
magnetic levitation force presented so far have taken different
assumptions based on different physical considerations. Most of
the models are based on the critical state model
\cite{bean62,bean64}, which has long been used to account for the
irreversible properties of type-II superconductors. The image
model \cite{yang92,yang94,yang96}treats the superconductor as a
pure diamagnet and the PM as a set of magnetic dipoles, if the
hysteretic behavior is not important. The system is described by
two magnetic dipoles, representing the PM and its mirror image,
oppositely magnetized and located the same distance below and
above the superconductor surfaces. The advantage of this model is
that analytical expressions can be obtained. However it cannot be
used to account for the dynamic stiffness.  The extension of the
image model to include another 'frozen' image of the PM has been
introduced to explain the dynamic stiffness of the system
\cite{kordyuk98}.

In a recent article by Navau and Sanchez \cite{navau98}, the
authors reviewed the model in the critical state. Early model
considered only extreme limits such as complete flux exclusion or
complete flux penetration, thus described the behavior of type-I
superconductors or type-II superconductors with very high critical
current \cite{yang92,hellman88}. This model was later extended to
describe the flux penetration
\cite{sanchez96,torng92,schonhuber94}. However, these models
assumed a superconducting sample small enough to consider the
magnetic-field-gradient constant along it. At the same time, the
demagnetization effect due to the finite dimensions of the
superconductor has been neglected. Navau and Sanchez
\cite{navau98} have accounted for the demagnetization by
introducing a demagnetization factor. However, it should be
pointed out that the use of demagnetization factor for
superconductors in the mixed state in which current is distributed
inside the bulk sample, is qualitatively invalid
\cite{daumling89,conner91}. Although the authors considered a
non-uniform field gradient along the sample, the radial magnetic
field generated by the PM has been neglected. While in real
experiments, the PM is usually smaller than the HTS, in this case
the assumption of neglecting the radial field may not be
applicable.

The models mentioned above can be used to account for some of the
levitation force features exhibited by experiments, however, they
are obviously oversimplified. First, the current and field
distribution in the HTS under a non-uniform magnetic field have
been neglected. For practical applications using the magnetic
levitation, the precise control of the interacting force between a
superconducting target and the applied non-uniform magnetic field
is of critical importance. It is hence necessary to thoroughly
understand the hysteretic force characteristics of a
superconductor immersed in such fields. From an analytical point
of view, a precise knowledge of current and field profiles is
essential to the force calculation of the magnetic levitation
force between the PM and the HTS. Second, the models are all based
on the critical state model, which neglects flux creep effects.
However, at the nitrogen temperature, which is the normal
operating temperature for the levitation force experiments, the
flux creep may play an important role on the levitation force.
Numerical analysis such as finite element method (FEM) has been
applied to calculate the magnetic levitation force, it is capable
of deriving the current and field profiles inside the HTS, but it
neglects the flux creep as well.

To better understand the design of the magnetic levitation system
between the PM and the HTS, a better model is required. In this
paper, taking all the above points into account, we develop a
model to describe a more realistic experimental configuration and
to account for the experimentally observed features of the
magnetic levitation force. The demagnetization effect is
considered by calculating the current and magnetization of finite
thickness superconductors, and current and field distributions in
the superconductor are derived. The flux creep effect is taken
into account by using a current-voltage law $E(J)=E_c(J/J_c)^n$.

The paper is structured as follows, In Sec.\ \ref{sec:modeling} we
will discuss the theoretical consideration for both the permanent
magnet and the superconductor in a non-uniform magnetic field,
including assumptions made and numerical considerations. The
results of the calculations, including the effects of the HTS
geometry and characteristics on the levitation force, the force
creep effects, the minor loops and the stiffnesses will be
presented and compared with experiments in the literature in
Sec.\ \ref{sec:redi}, and Sec.\ \ref{sec:conc}  are the concluding
remarks.

\section{\label{sec:modeling}Modeling}
\subsection{\label{pm}The Permanent magnet}

We consider a superconducting disk with radius $a$ and thickness
$2b$, levitated over a co-axial cylindrical permanent magnet with
radius $R_{\rm PM}$ and thickness $t_{\rm PM}$. The top surface
center of the PM is taken as the origin of the cylindrical
coordinate system $(\rho,\phi,z)$. Because of the axial symmetry
of the system, only the cross section of the system is considered,
with the axis direction $Z$ chosen as the symmetry axis of the SC
and the PM, and $\rho$ (radial direction) parallel to the surfaces
of the SC and the PM. For this configuration, the vector potential
of the PM has only one component along the $\phi$ direction and
can be derived by integrating the vector potential of a circular
current loop with radius $R_{\rm PM}$ along the thickness $t_{\rm
PM}$ as,
\begin{widetext}
\begin{equation}
A_\phi(\rho,z)=\frac{B_{\rm rem}}{2\pi}\int_0^\pi R_{\rm PM}\cos\phi\ln\frac{(z+t_{\rm PM})+\sqrt{R_{\rm PM}^2+\rho^2-2\rho R_{\rm PM}\cos\phi+(z+t_{\rm PM})^2}}{z+\sqrt{R_{\rm PM}^2+\rho^2-2\rho R_{\rm PM}\cos\phi+z^2}}d\phi,
\label{aphi}
\end{equation}
\end{widetext}
where $B_{\rm rem}$ is the remanent induction of the PM. The
radial field $B_\rho=-\partial A_\phi/\partial z$ can then be
written as,
\begin{widetext}
\begin{equation}
B_\rho(\rho,z)=\frac{B_{\rm rem}}{\pi}\sqrt{\frac{R_{\rm PM}}{\rho}}\sum_{i=0}^1\frac{(-1)^i}{k_i}\left[\left(1-\frac{1}{2}k_i^2\right)K(k_i)-E(k_i)\right]
\label{bar}
\end{equation}
\end{widetext}
where $K$ and $E$ are complete elliptic integrals of the first and
second kind, respectively. And
$$
k_i^2=\frac{4\rho R_{\rm PM}}{(R_{\rm PM}+\rho)^2+(z+it_{\rm PM})^2},\ \ \ \ \ \ \ \ \ i=0,1.
$$
The field on the axis $B_z=(1/\rho)\partial{(\rho
A_\phi)}/\partial \rho$ is,
\begin{widetext}
\begin{equation}
B_z(\rho,z)=\frac{B_{\rm PM}}{2\pi}\int_0^\pi\frac{\rho R_{\rm PM}\cos\phi-R_{\rm PM}^2\cos^2\phi}{R_{\rm PM}^2+\rho^2-2\rho R_{\rm PM}\cos\phi}\sum_{i=0}^1\frac{z+it_{\rm PM}}{\sqrt{R_{\rm PM}^2+\rho^2-2\rho R_{\rm PM}\cos\phi+(z+it_{\rm PM})^2}}d\phi+\frac{A_\phi}{\rho}
\label{bz}
\end{equation}
\end{widetext}

\subsection{The superconducting disk in a non-uniform magnetic field}

The response of superconducting disks and strips in an uniform
applied magnetic field has been extensively studied by Brandt
\cite{brandt98,brandt96}. We now consider the response of a
superconducting disk in the non-uniform magnetic field generated
by the PM [Eqs.~(\ref{bar}) and (\ref{bz})]. The calculation is
for the zero-field-cooled (ZFC) process, in which the
superconducting disk is cooled below its critical temperature
$T_c$ in zero applied field and then the non-uniform field is
applied to the disk. Screening current is generated inside the
superconducting disk. The central idea is to find the equation of
motion for the current density. Because of the axial symmetry, the
current density $J$ and the vector potential $A_J$ generated by
the current have only one component along the $\phi$ direction.
The total vector potential of the system is then $A=A_J+A_\phi$,
and the total magnetic field $B=\nabla\times A=\nabla\times A_J$.
We assume here the material law $\mathbf{B}=\mu_0\mathbf{H}$,
which is a good approximation when the flux density $B$ and the
critical sheet current $J_cb$ are larger than the lower critical
field $B_{c1}$ everywhere inside the superconducting disk. This
requirement is often satisfied in the magnetic levitation
measurement, normally operated at a relatively high temperature of
77 K. According to the Maxwell equation with gauge $\nabla\cdot
A_J=0$, we have $\mu_0J=\nabla\times B=\nabla\times\nabla\times
A=\nabla\times\nabla\times A_J=-\nabla^2A_J$. The solution of this
Laplace equation in cylindrical geometry can be written as

\begin{equation}
A_J(\rho, z)=-\mu_0\int_0^ad\rho^\prime\int_{-b}^bdz^\prime Q(r,r^\prime)J(r^\prime),
\label{aj}
\end{equation}
with $r=(\rho,z)$ and $r^\prime=(\rho^\prime,z^\prime)$. The integral kernel

\begin{equation}
Q(r,r')=f(\rho,\rho^\prime,z-z^\prime)+f(\rho,\rho^\prime,z+z^\prime),
\label{kernel}
\end{equation}
with
\begin{equation}
f(\rho,\rho^\prime,\eta)=\frac{-1}{\pi k}\sqrt{\frac{\rho^\prime}{\rho}}\left[\left(1-\frac{1}{2}k^2\right)K(k)-E(k)\right]
\end{equation}
$$
k^2=\frac{4\rho\rho^\prime}{(\rho+\rho^\prime)^2+\eta^2},
$$
$K$ and $E$ are the complete elliptic integrals of the first and
second kind, respectively.

Eq.~(\ref{kernel}) is obtained by integrating the 3D Green function
of the Laplace equation, $1/4\pi|r_3-r_3^\prime|$ with
$r_3=(x,y,z)$, over the angle $\phi=\arctan(y/x)$ \cite{brandt96}.
The total vector potential can then be written as
\begin{equation}
A(\rho,z)=-\mu_0\int_0^ad\rho^\prime\int_{-b}^{b}dz^\prime Q(r,r^\prime)J(r^\prime)+A_\phi(\rho,z)
\label{a}
\end{equation}

To obtain the desired equation of motion for the current density
$J(\rho,z,t)$, we express the induction law $\nabla\times \mathbf
E=-\dot{\mathbf B}=\nabla\times\dot{\mathbf A}$ in the form
$\mathbf E=-\dot{\mathbf A}$. Combining this expression with
Eq.~(\ref{a}), we have

\begin{equation}
E(\rho,z)=\mu_0\int_0^ad\rho^\prime\int_{-b}^bdz^\prime Q(r,r^\prime)\dot{J}(r^\prime)-\dot{A}_\phi(\rho,z)
\label{e}
\end{equation}

The equation of motion for the current density can be derived by
inverting Eq.~(\ref{e}) as
\begin{equation}
\dot{J}(\rho,z)=\frac{1}{\mu_0}\int_0^ad\rho^\prime\int_{-b}^bdz^\prime Q^{-1}(r,r^\prime)[E(r,r^\prime)+\dot{A}_\phi(\rho^\prime,z')],
\label{jdot}
\end{equation}
where $Q^{-1}$ is the reciprocal kernel defined by
\begin{equation}
\int_0^ad\rho^\prime\int_{-b}^bdz^\prime Q^{-1}(r,r^\prime)Q(r^\prime,r^{\prime\prime})=\delta(r-r^{\prime\prime})
\label{rq}
\end{equation}

It can be seen from Eq.~(\ref{jdot}) that the equation of motion
for the current density contains the reciprocal kernel, which
depends only on the geometry of the superconducting sample. Here
we consider a disk, but the kernel $Q$ can be changed to calculate
other sample geometries, such as strips, and even arbitrary sample
shapes \cite{brandt96}.  The equation also depends on the applied
magnetic field via $A_\phi$. Here we consider the field of a
permanent magnet given by Eqs.~(\ref{bar}) and (\ref{bz}), but one
may also use other type of magnetic field, such as the field
generated by a circular current loop or by dipoles. Besides these,
the equation also depends on the material law $E=E(J)$ of the
superconducting sample. Obviously any sufficiently sharply bent
$E(J)$ law may be used; in the following we choose the rather
general material law $E(J)=E_c(J/J_c)^n$, which yields the limits
of the critical state model for $n\to\infty$, the flux flow model
for $n=1$, and the flux creep model for $1<n<\infty$.
Eq.~(\ref{jdot}) is easily time integrated by starting with
$J(\rho,z,t=0)=0$ at time $t=0$ and then putting,
$J(\rho,z,t=t+dt)=J(\rho,z,t)+\dot{J}(\rho,z,t)dt$. The vector
potential can then be derived from Eq.~(\ref{a}) and the magnetic
field is $B=\nabla\times A$. For all the calculations in this
paper, we use a reduced unit of $E_c=a=\mu_0=1$. And the critical
current is assumed to be independent of the magnetic field.

\subsection{The levitation force and the hysteresis loop}

The PM approaches and recedes from the HTS as
$z=z_0-z_0\sin(\omega t)+z_{00}$, where $z_0+z_{00}$ is the
initial distance and $z_{00}$ is the minimum distance between the
PM and the HTS, $\omega$ represents the speed at which the PM
approaches and recedes from the HTS. Experimentally, uncertainty
will be caused when the PM touches the HTS, and therefore the
limit $z_{00} =0$ should be avoided. In this calculation we choose
$z_{00}/a=0.1$ as the minimum distance between the PM and the
superconducting disk.

As the current density and the magnetic field inside the
superconducting disk have been derived, the vertical levitation
force along the $z$-axis can be readily obtained as
\begin{equation}
F_z=2\pi\int_0^ad\rho\int_{-b}^bdz J(\rho,z)B_\rho(\rho,z)
\label{fz}
\end{equation}

As the applied magnetic field is non-uniform, in order to show the
magnetization hysteresis loop, we choose the axis field $B_{\rm
axis}=B_z(\rho=0,z)$ at the bottom surface center of the
superconducting disk to represent the strength of the applied
non-uniform field, with
\begin{equation}
B_{\rm axis}=\frac{B_{\rm rem}}{2}\left[\frac{z+t_{\rm PM}}{\sqrt{R_{\rm PM}^2+(z+t_{\rm PM})^2}}-\frac{z}{\sqrt{R_{\rm PM}^2+z^2}}\right]
\label{bz0}
\end{equation}
and the magnetic moment
\begin{equation}
m=2\pi\int_0^ad\rho\int_{-b}^bdz \rho^2J(\rho,z)
\label{m}
\end{equation}

\subsection{The minor force loop and the stiffness}

Because of the hysteretic behavior of the superconducting disk,
when the PM is moved away and back again by a small amount $\delta
z$ at a distance $z$, the $F_{\rm z}(z)$ curve follows a minor
force loop rather than the major force loop. For small $\delta z$,
the minor force loop is reversible, but beyond a certain $\delta
z$, hysteretic behavior in the minor force loop will be observed.
The stiffness at different distance $z$ can then be derived as
$K_z=-\partial F_z/\partial z$.

\begin{figure}[t]
\includegraphics*[bb=236 213 397 568, scale=1]{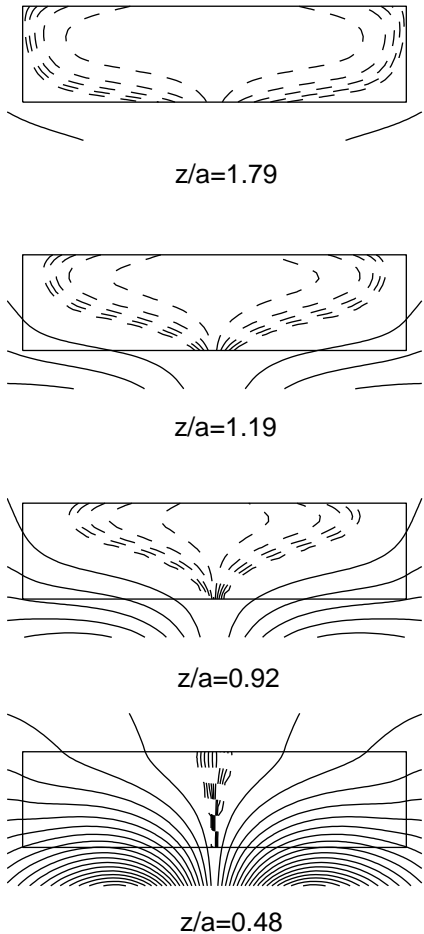}
\caption{Magnetic field lines when a PM approaches a
superconducting disk with $b/a=0.25$, $J_c=1$, and $\sigma=20$ at
distances $z/a=1.79, 1.19, 0.92, 0.48$. Dashed lines are the
contour lines of the current density inside the disk. The PM
approaches the disk as $z=z_0-z_0\sin(\omega t)$ with a frequency
of $\omega=0.1$.} \label{b}
\end{figure}

\begin{figure*}[t]
\includegraphics*[bb=53 206 552 612, scale=1]{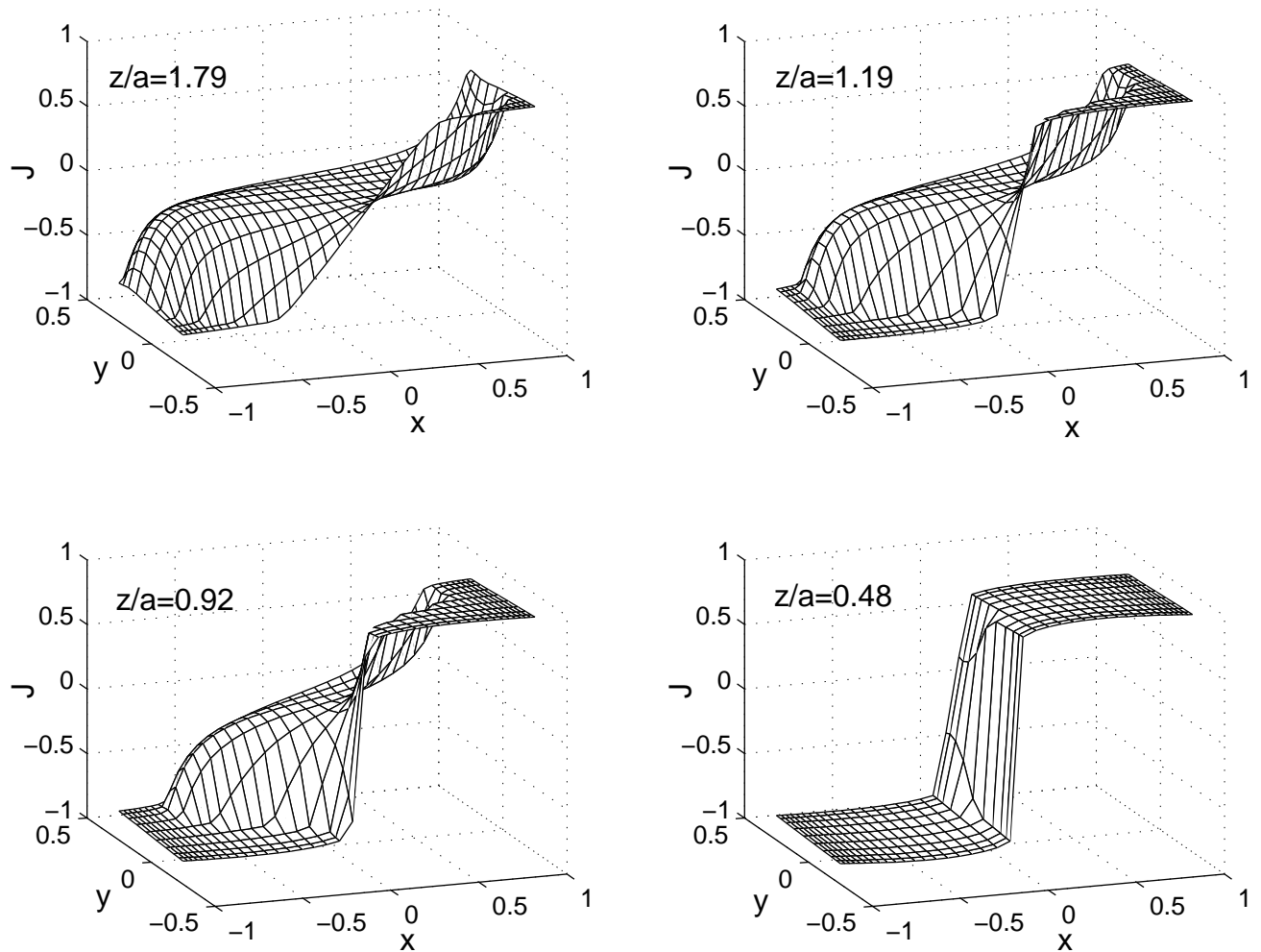}
\caption{Current profiles $J(x,y)$ for the same disk as in
Fig.~\ref{b}.} \label{j}
\end{figure*}

\section{\label{sec:redi}results and discussions}
\subsection{Current and field profiles in the HTS}

The magnetic flux lines when the PM is approaching the
superconducting disk with side ratio $b/a=0.25$ and $\sigma=20$,
$J_c=1$ are shown in Fig.~\ref{b} as solid lines at distances
$z/a=1.79$, $1.19$, $0.92$ and $0.48$. The dashed lines are the
contour lines of the current density inside the disk. Fig.~\ref{b}
shows the penetration of the non-uniform magnetic field generated
by the PM into the superconducting disk. Some features different
from those of the superconducting disk in uniform magnetic field
\cite{brandt98} can be clearly seen. Because the magnetic field is
stronger at the bottom of the disk (see $z/a=0.48$ of
Fig.~\ref{b}), the penetration starts from the bottom of the disk,
while the top surface is not penetrated. The resulting magnetic
field fronts inside the disk form an onion-shape rather than the
symmetrical lens-shape observed for homogeneous applied field.

Fig.~\ref{j} shows the profiles of the current density $J(x,y)$
corresponding to Fig.~\ref{b}. The current density first saturates
at the edges of the bottom surface, while the current density at
the edges of the top surface is smaller ($z/a=1.79$). As the PM is
moving closer, the saturation spreads both into the top surface
and into the middle of the disk ($z/a=1.19$ and $z/a=0.92$), until
it is saturated everywhere inside the disk ($z/a=0.48$). Because
we choose $\sigma=20$, the maximum current density is much smaller
than the critical current density $J_c=1$. Increasing $\sigma$
will result in a larger current density close to $J_c$.

\begin{figure}[t]
\includegraphics*[bb=236 213 397 568, scale=1]{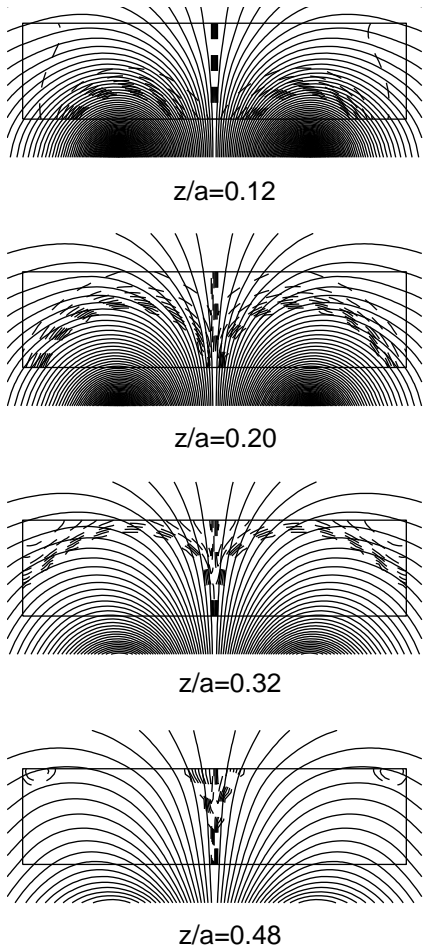}
\caption{Magnetic field lines when a PM is moving away from a
superconducting disk with $b/a=0.25$, $J_c=1$, and $\sigma=20$ at
distances $z/a=0.12, 0.20, 0.32, 0.48$. Dashed lines are the
contour lines of the current density inside the disk. The PM
recedes from the disk as $z=z_0-z_0\sin(\omega t)$ with a
frequency of $\omega=0.1$.} \label{ba}
\end{figure}

\begin{figure*}[t]
\includegraphics*[bb=52 206 552 612, scale=1]{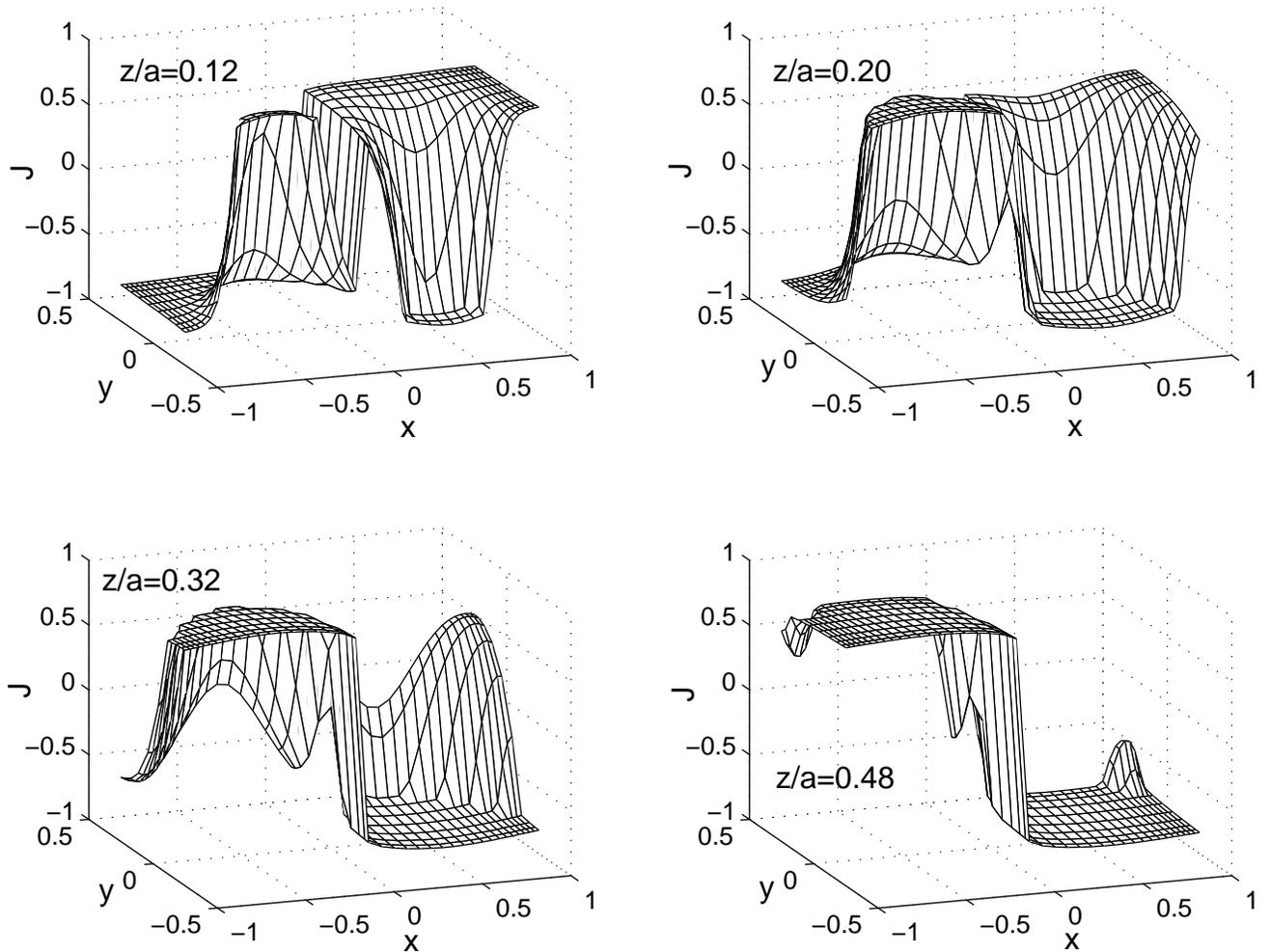}
\caption{Current profiles $J(x,y)$ for the same disk as in Fig.~\ref{ba}.}
\label{ja}
\end{figure*}

The magnetic field lines when the PM is moving away from the
superconducting disk with side ratio $b/a=0.25$ and $\sigma=20$,
$J_c=1$ are shown in Fig.~\ref{ba} as solid lines at distances
$z/a=0.12$, $0.20$, $0.32$ and $0.48$. The dashed lines are the
contour lines of the current density inside the disk. And
Fig.~\ref{ja} shows the corresponding profiles of the current
density $J(x,y)$. As the magnetic field is decreased, the current
density is reversed inside the disk. What is interesting is that
the reversion begins at the bottom surface with $x/a=\pm 0.5$,
where the magnetic field is strongest. As the PM is moving away
further, the reversion spreads onto the top surface and middle of
the disk (Fig.~\ref{ja}, $z/a=0.20$ and $z/a=0.32$), until the
current is completely reversed ($z/a>0.48$). Note that for a
superconducting disk immersed in a uniform magnetic field, the
reversion usually begins at the edge of the disk and spreads into
the center of the sample. As the levitation force is determined by the current density $J(x,y)$ and the magnetic field shown in Fig.~\ref{b} to Fig.\ref{ja}, the features shown in these figures will be reflected in the levitation force as will be discussed in the following.

The magnetic field lines and the profiles of the current density
$J(x,y)$ depend on the side ration $b/a$, $\sigma$ and critical
current density $J_c$ of the superconducting disk, as well as the
dimensions of the PM. Detailed results will be presented
elsewhere.

\subsection{Geometry of the HTS on $\mathbf{F_{\rm z}}$}

As we use reduced units of $E_c=a=\mu_0=1$ in this calculation,
the effect of the HTS geometry on the levitation is demonstrated
by calculating the levitation force for different thicknesses $2b$
of the superconducting disk. The results are shown in
Fig.~\ref{thickness}a, where $F_{\rm z}(z)$ curves at $\sigma=2$,
$\omega=0.1$ and $J_c=1$ for different thicknesses $2b$ are
plotted. The $F_{\rm z}$ curves show typical hysteretic behavior.
For larger side ratio $b/a$, the hysteretic loop is asymmetrical,
with the maximum repulsive force larger than the maximum
attractive force. However, as the side ratio is decreased, the
hysteretic force loop becomes more and more symmetrical. For
$b/a=0.05$, the curve is completely symmetrical. Experimentally,
symmetrical $F_{\rm z}$ versus $z$ curves have been observed in
YBCO thin film \cite{riise99}, while in MTG bulk samples, $F_{\rm
z}$ versus $z$ curves are usually asymmetrical. The reason for
this is easy to understand: For a thin sample, the magnetic moment
is saturated when the PM is close to the sample. When the PM is
moving away from the sample, only a little decrease in the applied
field saturates the magnetic moment in the reverse direction,
results in a symmetrical magnetization hysteresis loop (see
Fig.~\ref{thickness}b, where the magnetization hysteresis loops
corresponding to Fig.~\ref{thickness}a are plotted), and therefore
a symmetrical force loop. While for a bulk sample the magnetic
moment saturates when the PM is close to the sample, it is never
saturated in the reverse direction when the PM is moving far away
from the disk (see the magnetization hysteresis loops for $b/a\ge
0.8$ in Fig.~\ref{thickness}b), resulting in an asymmetric
hysteresis loop of the magnetization and therefore an asymmetric
loop of the force versus distance.

Another interesting feature shown in Fig.~\ref{thickness} is that
a maximum is found in the repulsive force when the PM is at some
distance away from the HTS, rather than at the minimum distance.
Experimentally this maximum has been observed in thin films
\cite{riise99}, while it was hardly observed in MTG samples. Riise
et al. accredited this maximum to the dimensions of the PM, which
tends to vanish with larger $t_{\rm PM}$ and smaller $R_{\rm PM}$.
And the disappearance of the maximum in MTG sample was explained
noting that bulk samples are less sensitive to a non-uniform field
than films \cite{riise99}. Another explanation comes from Sanchez
and Navau \cite{sanchez96}, who claimed that this maximum is a
result of the minimum in the derivative of the field produced by
the PM, based on a constant-field-gradient model \cite{navau98}.
They concluded that the maximum depends crucially on the side
ratio $b/a$, the larger the side ratio $b/a$, the closer the
maximum shifts to zero PM-HTS separation. When $b/a$ is sufficiently large, the
maximum may not exist. However, as can be seen from
Fig.~\ref{thickness}a, the maximum is independent of the side
ratio $b/a$. The peaks are observed to be at the same distance for
all side ratios and even for $b \gg a$ not shown here. In a
separate calculation, we have found that this maximum is
independent of the dimensions of the PM. We conclude that the peak
arises from the intrinsic properties of the HTS. As will be seen
from figures below, the maximum depends on the creep exponent
$\sigma$. The larger $\sigma$, the closer the maximum shifts to
zero PM-HTS separation. When $\sigma$ is sufficiently large, the
maximum may not be observed any more.

Because $\sigma=n-1$, and $n$ can be related to the pinning
potential of the sample as $n=U_0(T,B)/k_BT$
\cite{brandt96,brandt98}, smaller $\sigma$ means lower pinning
potential or higher temperature. Both MTG YBCO and YBCO thin film
have a relatively high pinning potential, however experimentally
when a PM approaches a thin film, it may increase the local
temperature on the film, resulting in a lower $\sigma$, and
therefore in the maximum. On the other hand, although the PM may
increase the temperature of the surface of the bulk sample, the
currents flow in a much larger volume and the levitation force is
determined by the bulk properties, therefore the maximum may not
be observed in MTG bulk samples.

In the inset of Fig.~\ref{thickness}a, we show the maximum
repulsive force as a function of the side ratio $b/a$. It can be
seen from the figure that for small side ratio $b/a$, the maximum
repulsive force increases linearly with $b/a$, but saturates as
$b/a$ is further increased. Technically, a superconducting disk
with diameter $2a$ approximately equal to the thickness may be
optimum for magnetic levitation, further increase of the
thickness will not enhance the levitation force. This calculated
result is consistent with the experimental observations
\cite{unsworth93}.

\subsection{Characteristics of the HTS on $\mathbf{F_{\rm z}}$}

\begin{figure}[t]
\includegraphics*[bb=115 280 480 560, scale=0.65]{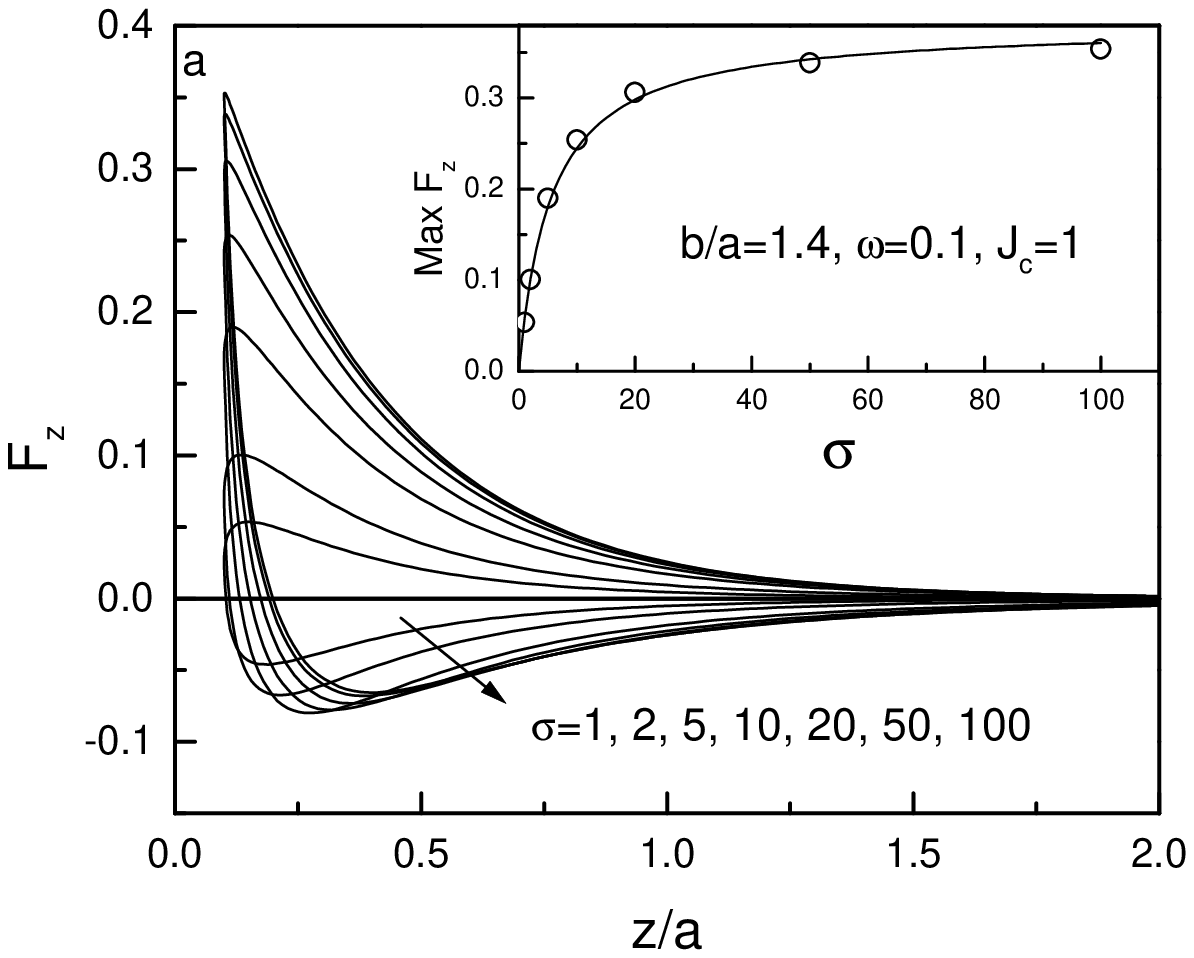}
\includegraphics*[bb=115 280 480 560, scale=0.65]{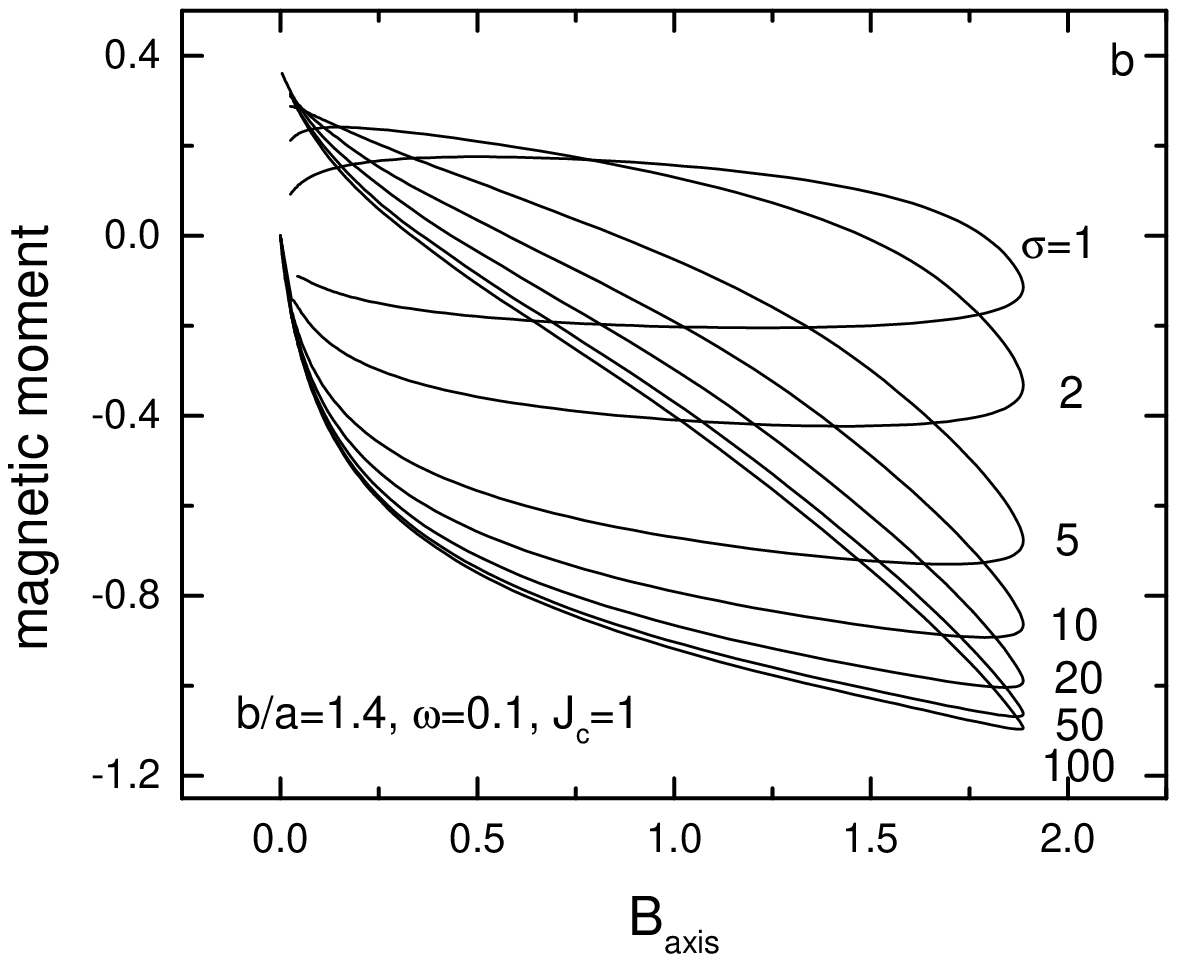}
\caption{(a) The vertical magnetic levitation force $F_{\rm z}$
versus the distance $z$ at $b/a=1.4$, $\omega=0.1$ and $J_c=1$ for
different pinning potentials of the superconducting disk. Inset
shows the maximum repulsive force as a function of $\sigma$, the
solid line is a guide for the eyes only. (b) The corresponding
magnetization hysteresis loops.} \label{sigma}
\end{figure}

Characteristics of the HTS in this calculation are represented by
two parameters. One is the creep exponent $\sigma$ related to the
pinning potential by $n=\sigma+1=U_0(T,H)/k_BT$; another one is the
critical current density $J_c$ representing the pinning strength.
The effect of $\sigma$ on the magnetic levitation force is shown
in Fig.~\ref{sigma}a, where the vertical magnetic levitation force
$F_{\rm z}$ versus the distance $z$ at $b/a=1.4$, $\omega=0.1$ and
$J_c=1$ for different pinning potentials of the superconducting
disk are plotted. It can be seen from Fig.~\ref{sigma}a that as
$\sigma$ is increased, the hysteretic force loop increases and the
maximum repulsive force shifts to smaller PM-HTS separation. When
$\sigma$ is sufficiently large, the peak is not observed, as
discussed in the above section (the maximum at $\sigma=100$
corresponds to minimum HM-HTS separation). On the contrary, the
maximum in the attractive force shifts to larger PM-HTS separation
with increasing $\sigma$. The maximum attractive force increases
with $\sigma$, however, when $\sigma$ is larger than 5, it
decreases again and saturates at higher $\sigma$. As opposed to
this, the maximum repulsive force increases monotonically with
$\sigma$ and reaches a saturation value at large $\sigma$, as
shown in the inset of Fig.~\ref{sigma}a.

\begin{figure}[t]
\includegraphics*[bb=115 280 480 560, scale=0.65]{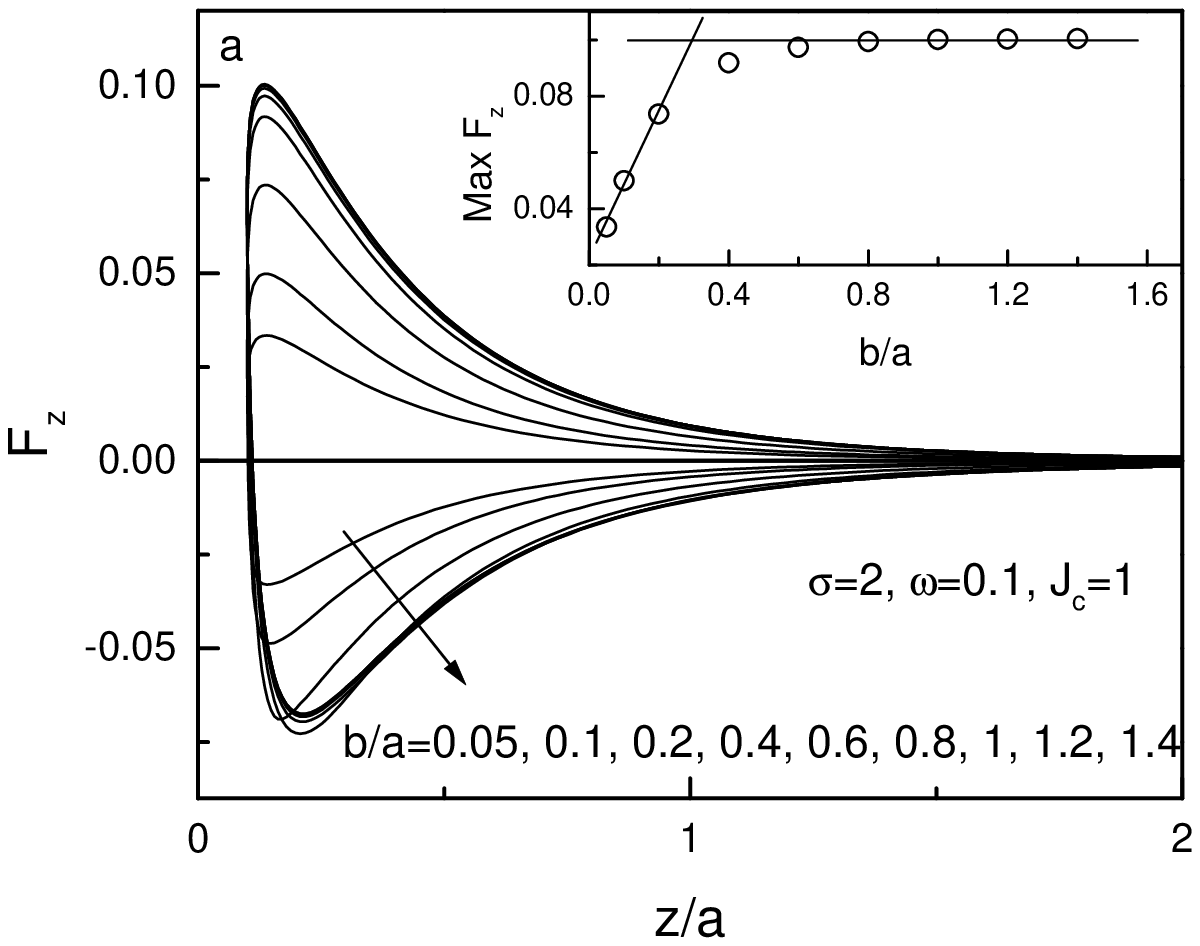}
\includegraphics*[bb=115 280 480 560, scale=0.65]{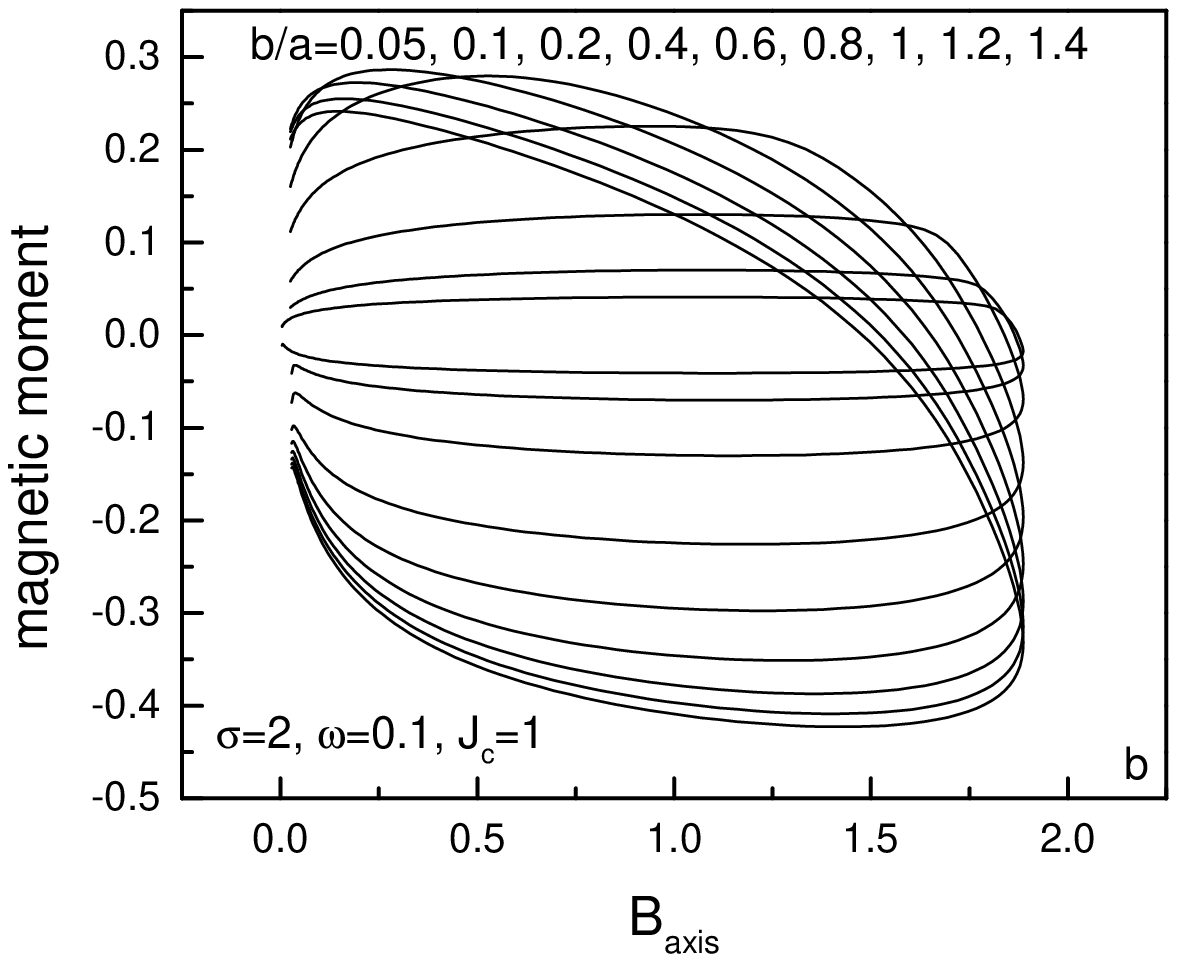}
\caption{(a) The vertical magnetic levitation force $F_{\rm z}$
versus the distance $z$ at $\sigma=2$, $\omega=0.1$ and $J_c=1$
for different thicknesses of the superconducting disk. Inset shows
the maximum repulsive force as a function of the thickness, solid
lines are guide for the eyes only. (b) The corresponding
magnetization hysteresis loops.} \label{thickness}
\end{figure}

Fig.~\ref{sigma}b shows the corresponding magnetization hysteresis
loops. For $\sigma=1$, the pinning potential is very small,
because of the relaxation effects, the applied magnetic field can
penetrate deeper into the sample, similar to the case of small
side ratio $b/a$ shown in Fig.~\ref{thickness}. The magnetization
hysteresis loop and therefore the force loop are symmetric.
Increasing the pinning potential (larger $\sigma$) results in
larger hysteresis loops of the magnetization. However, further
increase of  $\sigma$ leads to a reversible response and to a
smaller magnetization hysteresis loop. Interestingly, the peak in
the maximum attractive force corresponds to the maximum hysteresis of the magnetization (here $\sigma=5$) as can be seen from Fig.~\ref{sigma}, which is also
the case in Fig.~\ref{thickness}.

The sharp decrease of the levitation force from repulsive to
attractive results from the non-uniform magnetic field generated
by the PM and from the relaxation of the current in the HTS.
According to Eq.~(\ref{fz}), the vertical levitation force is
determined by the radial magnetic field for the PM and the current
density of the HTS. The radial magnetic field is strongest close
to $x/a=\pm R_{\rm PM}$ at the bottom of the disk [see
Eq.~(\ref{bar})]. Therefore, the radial magnetic field and the
current close to $x/a=\pm R_{\rm PM}$ on the bottom of the disk
contribute more to the levitation force. When the applied field is
decreased by a small amount, the current close to $x/a=\pm R_{\rm
PM}$ on the bottom of the disk is reversed to positive first, although the
total current is still negative (the magnetic moment is negative),
the levitation force decreases sharply to a negative value,
because the attractive force results from the current close to
$x/a=\pm R_{\rm PM}$ at the bottom of the disk is much larger than
the repulsive force resulting from the negative current in the HTS.
For a small $\sigma$, because of relaxation, the current is small, and only a small decrease
in the applied magnetic field may lead to a complete reversal of
the current, see Fig.~\ref{sigma}, therefore the maximum attractive force is closer to zero
PM-HTS separation. When $\sigma$ is increased, the current is
larger, more decrease of the applied magnetic field is
needed to reverse the current. Therefore the maximum attractive
force is larger and shifts to larger PM-HTS separation. Further
increasing $\sigma$ shifts the maximum attractive force to larger
PM-HTS separation, but the maximum attractive force is smaller.
This is because much more decrease of the
applied magnetic field is needed to reverse the larger current
when $\sigma$ is increased, although the current is larger but the
radial field is smaller, resulting in a smaller maximum attractive
force observed in Fig.~\ref{sigma}a.

\begin{figure}[t]
\includegraphics*[bb=115 280 480 560, scale=0.65]{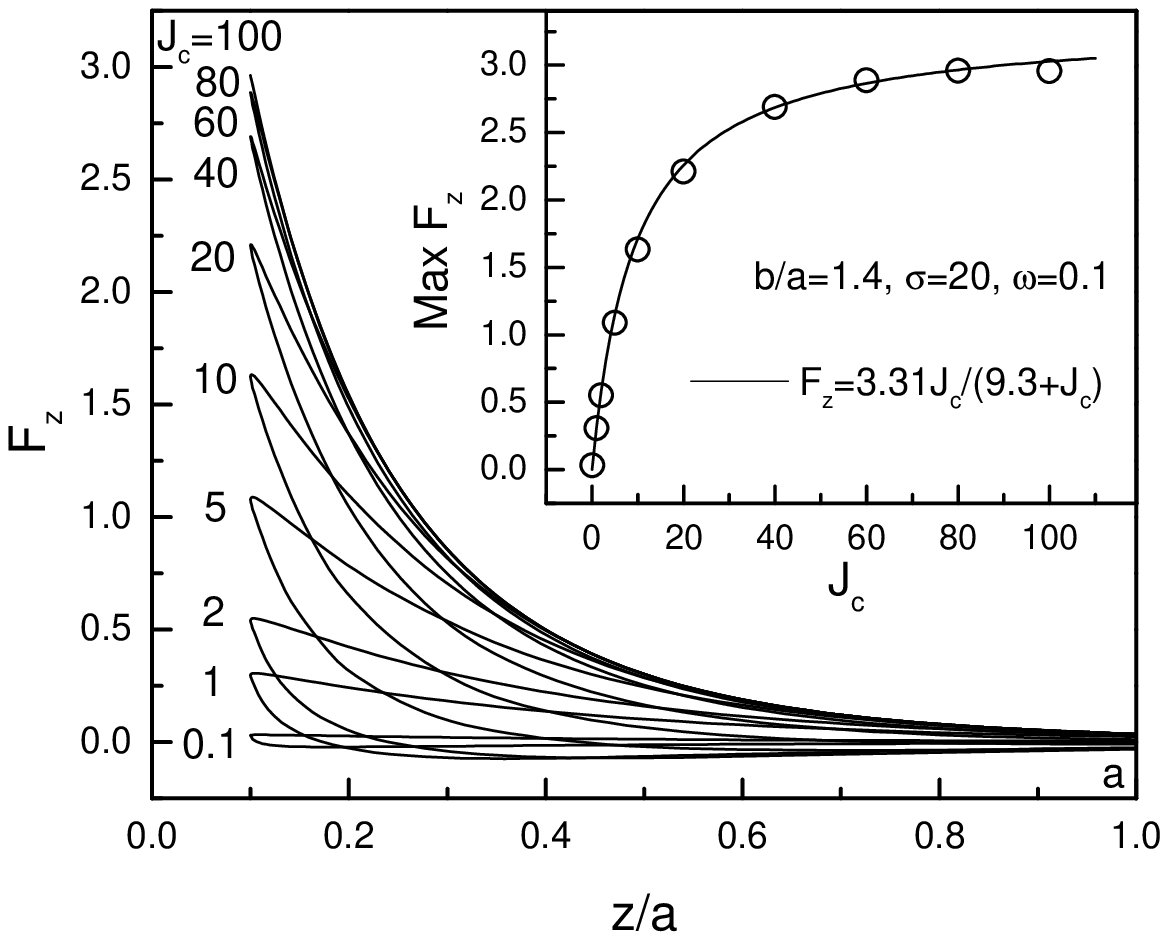}
\includegraphics*[bb=115 280 480 560, scale=0.65]{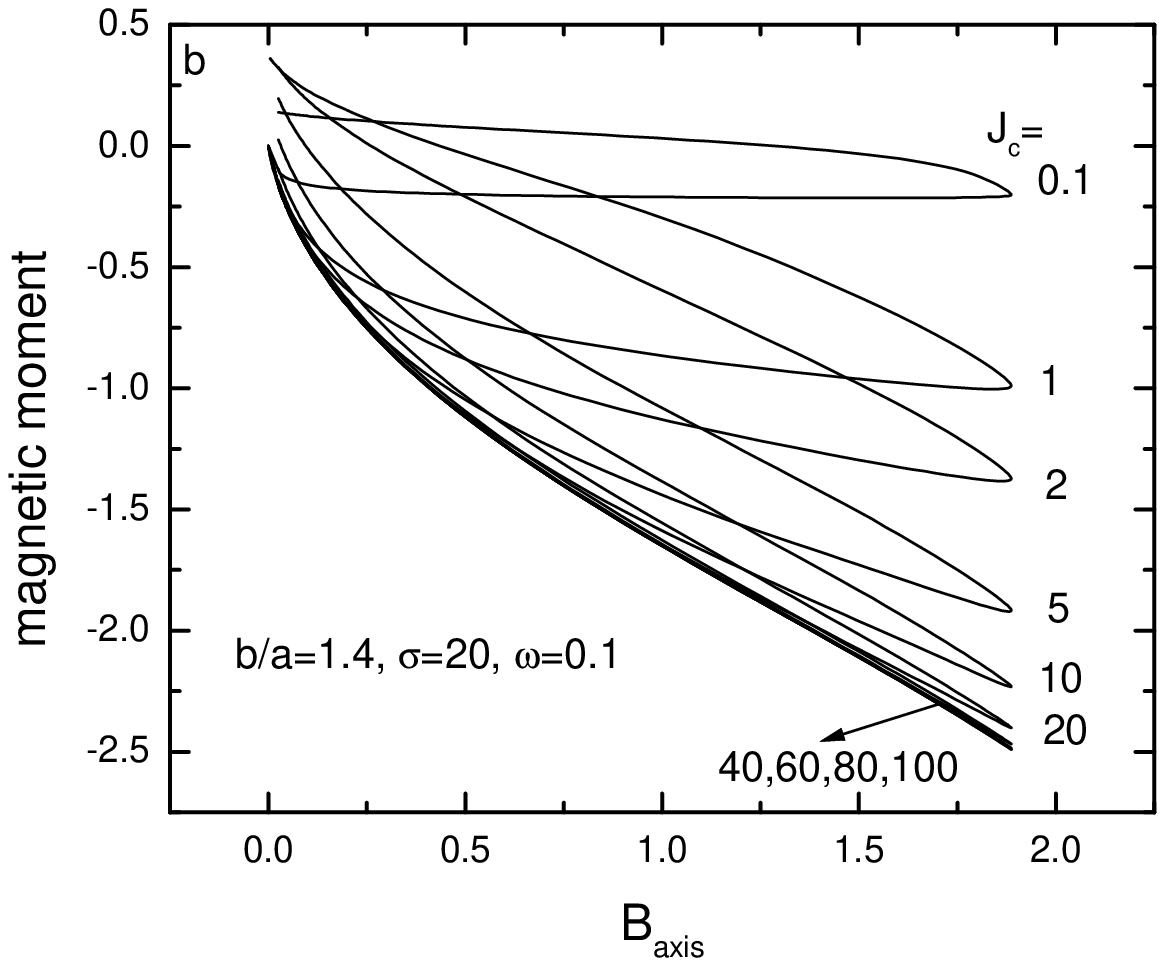}
\caption{(a)The vertical magnetic levitation force $F_{\rm z}$
versus the distance $z$ at $b/a=1.4$, $\sigma=20$ and $\omega=0.1$
for different critical current densities of the superconducting
disk. Inset shows the maximum repulsive force as a function of
$J_c$, the solid line is a fit with $F_{\rm z}=3.31J_c/(9.3+J_c)$.
(b)The corresponding magnetization hysteresis loops.} \label{jc}
\end{figure}

Another characteristics of the HTS which drastically influences
the magnetic levitation force is the critical current density of
the HTS. The calculated results of $F_{\rm z}$ versus $z$ curves
at $b/a=1.4$, $\sigma=20$ and $\omega=0.1$ for different critical
current densities $J_c$ are plotted in Fig.~\ref{jc}a. The
corresponding magnetization hysteresis loops are shown in
Fig.~\ref{jc}b. For small critical current densities ($J_c\le
10$), the calculated results are typical $F_{\rm z}$ versus $z$
curves, which show both the repulsive force branch and the
attractive force branch. The attractive force branches of the
curves are hidden by the large vertical axis scale. However, when
$J_c$ is larger than $20$, no attractive force is observed. When
$J_c$ is larger than $60$, the $F_{\rm z}$ versus $z$ curves are
almost reversible ($\sigma=80,100$). This result can also be seen
from the magnetization hysteresis loops, as shown in
Fig.~\ref{jc}b. Increasing the critical current density leads to
larger hysteresis loop and then to reversible magnetization
hysteresis loops. Experimentally, $F_{\rm z}$ versus $z$ curves with very
small hysteresis width has been observed in a MTG sample
\cite{hull99}, indicating a very high critical current density in
this sample.

Based on the critical state model, the constant-field-gradient
model \cite{sanchez96} predicted that the maximum repulsive force
depends linearly on the critical current density. In order to
study this dependence, we plot in the inset of Fig.~\ref{jc}a the
dependence of the maximum repulsive force as a function of the
critical current density shown as open circles. It is obvious that
the dependence is a nonlinear function. The linear dependence is
observed only at very low critical current density $J_c<5$. A
fitting to the obtained data results in the dependence
\begin{equation}
F_z=\frac{3.3J_c}{9.3+J_c},
\label{fzjc}
\end{equation}
shown as solid line in the inset. From Eq.~(\ref{fzjc}) we can see
that the maximum repulsive force saturates at high critical
current density.

Experimentally it is very difficult to get systematic results as
shown in Fig.~\ref{sigma} and Fig.~\ref{jc}, however,
Fig.~\ref{sigma} and Fig.~\ref{jc} can be used as a quick
reference for testing the properties of the sample under
investigation. By comparing the experimental results with
Fig.~\ref{sigma} and Fig.~\ref{jc}, one may have an idea about the
pinning potential and critical current of the sample.

\subsection{Force creep}

It has been well known that the current density in HTS experiences
relaxation (decay with time), which has been extensively studied
both theoretically and experimentally in terms of the
magnetization. Because the vertical magnetic levitation force is
determined by the current density in the HTS and radial magnetic
field generated by the PM, it is expected that the levitation
force may exhibit relaxation, resulting in a time dependent
levitation force in experiments. However, reports on this dynamic
behavior are very few \cite{riise92,moon90}. On the other hand, no
models so far have taken this relaxation behavior of the magnetic
levitation force into account. This is quite surprising
considering the operating temperature of the levitation system
between a HTS and a PM, which is normally at the relatively high
temperature of 77K, where the relaxation rate is usually quite
high. It is also surprising if one considers the effects of the
force decay on the design of any practical applications invoking
PM-HTS levitation.

\begin{figure}[t]
\includegraphics*[bb=115 280 480 560, scale=0.65]{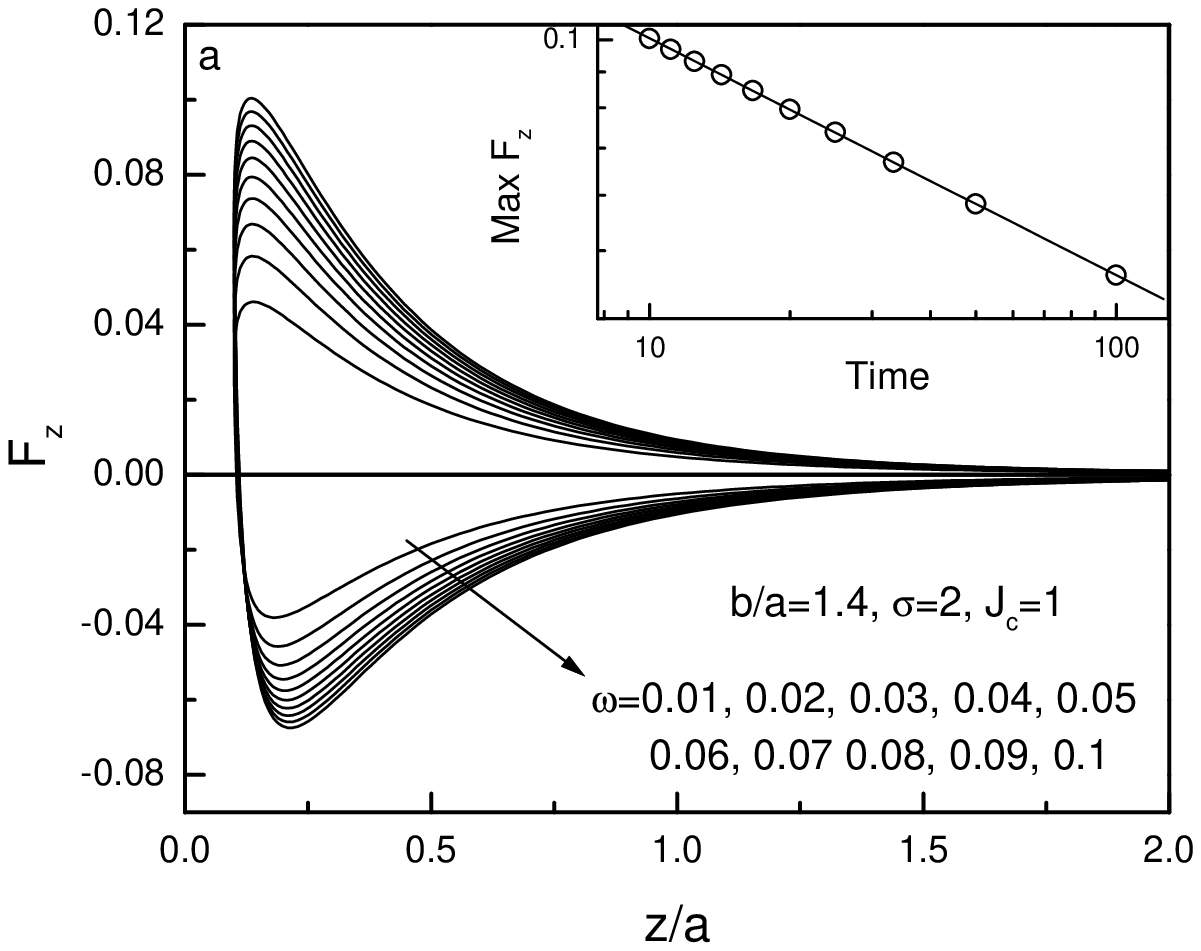}
\includegraphics*[bb=115 280 480 560, scale=0.65]{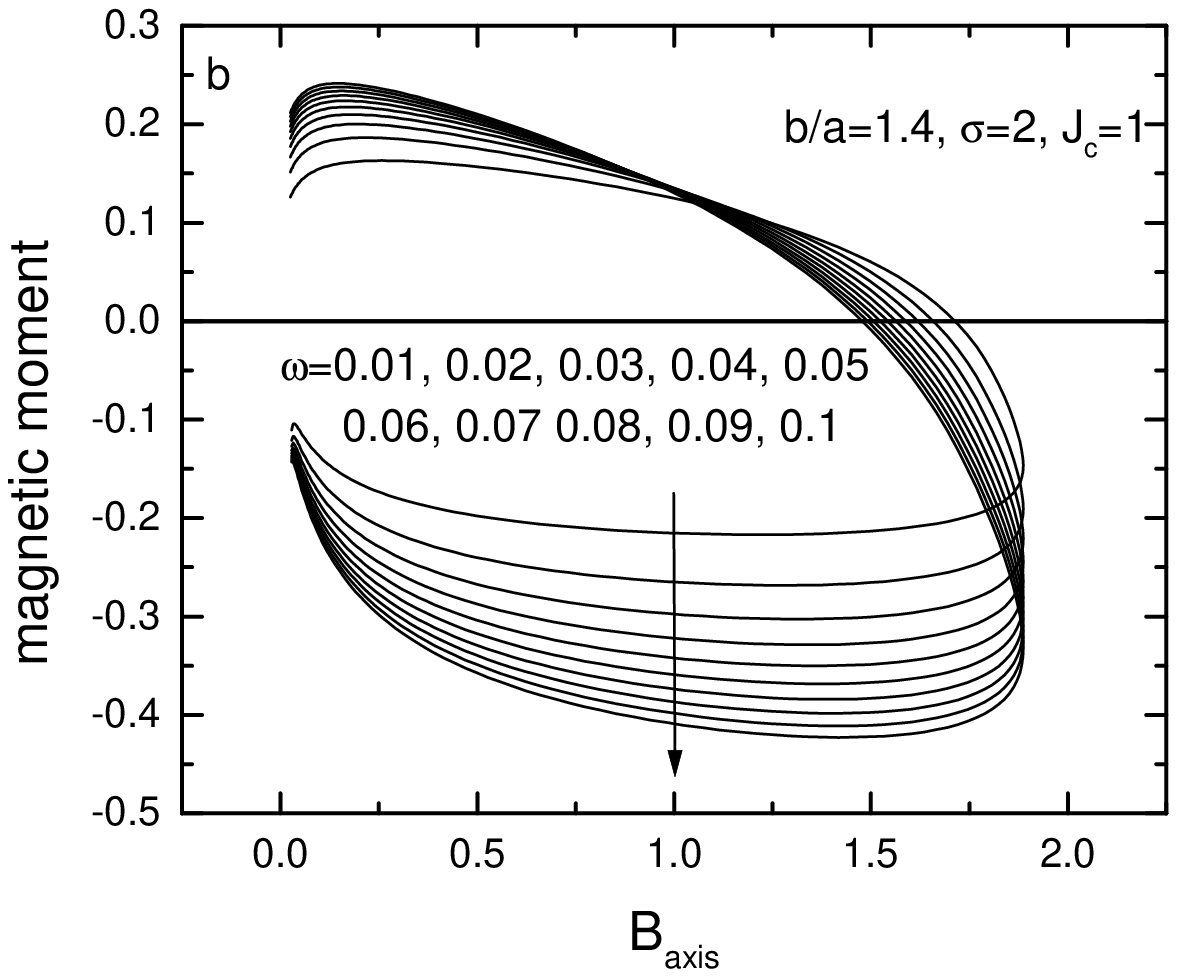}
\caption{(a)The vertical magnetic levitation force $F_{\rm z}$ versus the distance $z$ at $b/a=1.4$, $\sigma=2$ and $J_c=1$ for different frequencies at which the PM approaches and recedes from the HTS. Inset shows the relaxation of the levitation force. (b)The corresponding magnetization hysteresis loops.}
\label{relax}
\end{figure}

In this calculation, the force relaxation is taken into account by
using a voltage-current law $E(J)=E_c(J/J_c)^n$. With
$1<n<\infty$, this material law describes the flux creep in terms
of the magnetization inside the superconducting disk. In
Fig.~\ref{relax} we show the $F_{\rm z}$ versus $z$ curves at
$b/a=1.4$, $\sigma=2$ and $J_c=1$ for different frequencies
$\omega$ at which the PM approaches and recedes from the HTS. It
is obvious that the levitation force depends on the speed at which
the PM approaches and recedes from the HTS; larger speed leads to
a larger force loop. On the contrary, previous models based on the
critical state model assume that the levitation force is
independent of the speed at which the PM approaches and recedes
from the HTS \cite{hull99}.

In the inset of Fig.~\ref{relax}a, we show the maximum repulsive
force as a function of the time $t=1/\omega$ as open circles in a
double logarithmic plot. The linear dependence is obvious. Fitting
to the data gives a $F_{\rm z}\propto t^{-m}$ dependence of
$F_{\rm z}$ on time. This result actually arises from the
thermally activated flux motion in the HTS, because the voltage
current law $E(J)=E_c(J/J_c)^n$ is actually a result of the
current dependence of the activation energy $U(J)=U_0\log(J_c/J)$;
this can be shown by using the Arrhenius law
$E=Bv=Bv_0\exp[-U(J)/k_BT]$. With this activation energy, the
relaxation of the current density can be derived as $J\propto
t^{-m}$, leading to the force relaxation shown in the inset of
Fig.~\ref{relax}a. Experimentally, the force creep has been
observed to be logarithmic in time \cite{riise92} within a narrow
time window. It is expected that a nonlogarithmic force creep will
be observed if an extended time window is measured.

The force relaxation can also be clearly seen from
Fig.~\ref{relax}b, where the magnetization hysteresis loops
corresponding to Fig.~\ref{relax}a are plotted. The magnetization
hysteresis loops increase with increasing frequency. This behavior
is similar to the so-called dynamic magnetic moment relaxation
measurement, in which the DC magnetization hysteresis loops are
measured at different sweep rates of the applied magnetic moment.
The magnetization hysteresis loop increases with increasing sweep
rate. It has been proven that the dynamic magnetic moment
relaxation is equivalent to the normal relaxation measurement, in
which the magnetic moment is recorded as a function of time at
fixed temperature and magnetic field,  when studying the
relaxation of HTS \cite{schnack92,jirsa93}. Therefore, by
measuring the vertical levitation force $F_{\rm z}$ versus $z$ at
different frequencies with which the PM approaches and recedes
away from the HTS, one can study the relaxation of the force as
well as the current density of the HTS.

\subsection{Minor force loops and stiffness}

\begin{figure}[t]
\includegraphics*[bb=115 280 480 560, scale=0.65]{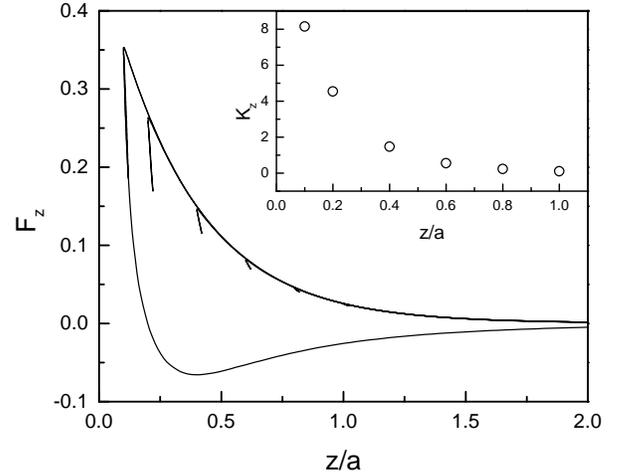}
\caption{(a)The vertical magnetic levitation force $F_{\rm z}$
versus the distance $z$ at $b/a=1.4$, $\sigma=100$, $\omega=0.1$
and $J_c=1$. The minor force loops at different distances $z$ are
shown with a amplitude $\delta z=0.02a$. Inset shows the stiffness
of the system.} \label{stiff}
\end{figure}

One of the most important parameters used to characterize the
magnetic levitation system using a PM and a HTS, is the magnetic
stiffness defined as $K_{\rm z}=-\partial F_{\rm z}/\partial z$,
which represents the spring constant associated with vibrational
motion of a levitation system. The magnetic stiffness can be
determined by measuring or computing minor force loops. The
calculated minor force loops at different PM-HTS separations for
$b/a=1.4$, $\sigma=100$, $\omega=0.1$ and $J_c=1$ are shown in
Fig.~\ref{stiff}. Here we use an amplitude of $\delta z=0.02a$,
resulting in reversible minor force loops. Higher amplitude will
lead to hysteretic behavior not shown here. The calculated
vertical stiffness is plotted in the inset as a function of the
PM-HTS distance $z$. It can be seen from Fig.~\ref{stiff} that
$K_{\rm z}>0$, indicating a stable levitation between a PM and a
HTS. $K_{\rm z}$ decreases with increasing PM-HTS separation $z$.
$K_{\rm z}$ may also depend on other parameters, such as $J_c$,
$\omega$, $\sigma$ and the shapes of both the PM and the HTS. All
these cases are easily computed by our numerical method. Detailed
results will be presented elsewhere.

\section{\label{sec:conc}conclusions}
The current density and magnetic field profiles of a
superconducting disk (SC) with radius $a$ and thickness $2b$
immersed in the non-uniform magnetic field generated by a
permanent magnet (PM) are calculated from first
principles for the superconductor. From the derived current density, the magnetic
levitation force between the SC and the PM has been derived by
assuming a voltage-current law $E(J)=E_c(J/J_c)^n$ and a material
law $\mathbf{B}=\mu_0\mathbf{H}$. The geometry and characteristics
of the SC drastically influence the vertical levitation force
$F_{\rm z}$. $F_{\rm z}$ depends non-linearly on the critical
current density $j_c$ of the SC: $F_z=3.31J_c/(9.3+J_c)$. For thin
samples, $F_{\rm z}$ depends linearly on the thickness of the
sample, but beyond a certain thickness, $F_{\rm z}$ is mostly
independent of the thickness. The flux creep also plays an
important role on the magnetic levitation force, which is observed
to be non-logarithmic in time. The stiffness for the SC and PM
system has also been derived. The calculated results reproduce
many of the features for the magnetic levitation between a SC and
a PM exhibited by experiments.

\section{acknowledgment}
The authors would like to thank the Australian Research Council
for financial support.

\end{document}